\documentclass[aps,pra,twocolumn,10pt,amsmath,superscriptaddress,amssymb,longbibliography]{revtex4-2}

\usepackage{epsfig}
\usepackage{epstopdf}
\usepackage{dcolumn}
\usepackage{amsbsy}
\usepackage{bm}
\usepackage{color,CJK}
\usepackage{amsmath,amssymb,amsfonts}
\usepackage{appendix}
\usepackage{graphicx}
\usepackage{algorithm}
\usepackage{enumerate}
\usepackage{makeidx}
\usepackage{braket}
\usepackage[usenames,dvipsnames]{xcolor}
\usepackage[driverfallback=dvipdfmx,colorlinks,linkcolor=blue,citecolor=blue,urlcolor=blue]{hyperref}
\usepackage[all]{hypcap}

\begin{document}
	
\title{Experimental study of matter-wave four-wave mixing in $^{39}$K Bose-Einstein condensates with tunable interaction}
	
\author{Yue Zhang }
\affiliation{State Key Laboratory of Quantum Optics Technologies and Devices, Institute of Opto-electronics, \\
\makebox[0.8\linewidth][c]{Collaborative Innovation Center of Extreme Optics, Shanxi University, Taiyuan, Shanxi 030006, China}}

\author{Liangchao Chen}
\email[Contact author: ]{chenlchao87@sxu.edu.cn}
\affiliation{State Key Laboratory of Quantum Optics Technologies and Devices, Institute of Opto-electronics, \\
\makebox[0.8\linewidth][c]{Collaborative Innovation Center of Extreme Optics, Shanxi University, Taiyuan, Shanxi 030006, China}}
\affiliation{Hefei National Laboratory, Hefei, Anhui 230088, China}
	
\author{Zekui Wang}
\affiliation{State Key Laboratory of Quantum Optics Technologies and Devices, Institute of Opto-electronics, \\
\makebox[0.8\linewidth][c]{Collaborative Innovation Center of Extreme Optics, Shanxi University, Taiyuan, Shanxi 030006, China}}
	
\author{Yazhou Wang}
\affiliation{State Key Laboratory of Quantum Optics Technologies and Devices, Institute of Opto-electronics, \\
\makebox[0.8\linewidth][c]{Collaborative Innovation Center of Extreme Optics, Shanxi University, Taiyuan, Shanxi 030006, China}}

\author{Pengjun Wang}
\affiliation{State Key Laboratory of Quantum Optics Technologies and Devices, Institute of Opto-electronics, \\
\makebox[0.8\linewidth][c]{Collaborative Innovation Center of Extreme Optics, Shanxi University, Taiyuan, Shanxi 030006, China}}
\affiliation{Hefei National Laboratory, Hefei, Anhui 230088, China}
	
\author{Lianghui Huang}
\affiliation{State Key Laboratory of Quantum Optics Technologies and Devices, Institute of Opto-electronics, \\
\makebox[0.8\linewidth][c]{Collaborative Innovation Center of Extreme Optics, Shanxi University, Taiyuan, Shanxi 030006, China}}
\affiliation{Hefei National Laboratory, Hefei, Anhui 230088, China}
	
\author{Zengming Meng}
\affiliation{State Key Laboratory of Quantum Optics Technologies and Devices, Institute of Opto-electronics, \\
\makebox[0.8\linewidth][c]{Collaborative Innovation Center of Extreme Optics, Shanxi University, Taiyuan, Shanxi 030006, China}}
\affiliation{Hefei National Laboratory, Hefei, Anhui 230088, China}
	
\author{Zhuxiong Ye}
\affiliation{State Key Laboratory of Quantum Optics Technologies and Devices, Institute of Opto-electronics, \\
\makebox[0.8\linewidth][c]{Collaborative Innovation Center of Extreme Optics, Shanxi University, Taiyuan, Shanxi 030006, China}}
	
\author{Wei Han}
\affiliation{State Key Laboratory of Quantum Optics Technologies and Devices, Institute of Opto-electronics, \\
\makebox[0.8\linewidth][c]{Collaborative Innovation Center of Extreme Optics, Shanxi University, Taiyuan, Shanxi 030006, China}}
\affiliation{Hefei National Laboratory, Hefei, Anhui 230088, China}
	
\author{Jing Zhang}
\email[Contact author: ]{jzhang74@sxu.edu.cn}
\affiliation{State Key Laboratory of Quantum Optics Technologies and Devices, Institute of Opto-electronics, \\
\makebox[0.8\linewidth][c]{Collaborative Innovation Center of Extreme Optics, Shanxi University, Taiyuan, Shanxi 030006, China}}
\affiliation{Hefei National Laboratory, Hefei, Anhui 230088, China}
	
\date{\today }
	
\begin{abstract}
We experimentally investigate four-wave mixing (FWM) of matter waves in two geometric configurations in $^{39}$K Bose-Einstein condensates with the atomic interaction tuned via Feshbach resonances. For one configuration with the single-spin component, the FWM yield increases with a larger scattering length. For the two-spin component configuration, we specifically investigate FWM in both the droplet and gas parameter regimes. We find that the FWM yield reaches its maximum near the critical parameter region between the gas and droplet phases.  Our research can help to optimize the FWM yield for matter-wave amplification and entangled atom pair generation, making it conducive to applications in quantum information processing and precision measurement.
\end{abstract}
\maketitle

\section{INTRODUCTION}
	
Atoms and photons share many similarities in their particle-like and wave-like behaviors. Photons can undergo a nonlinear four-wave mixing (FWM) process, wherein the input of three photons into the medium coherently generates a fourth photon in distinct modes \cite{RevModPhys.54.685, NJP123023, vaughan2007coherently, qiu2020highly}. Optical FWM enables the generation of nonclassical photonic states, allowing it to play a vital role in diverse areas of science and technology \cite{podhora2017nonclassical, morris2014photon, zatti2023generation, dmitriev2017quantum}. Similar to photons, atoms in coherent matter waves can also undergo the FWM process, that is, the interplay of three matter-wave packets in different momentum modes will generate a new matter-wave packet. The FWM of matter waves occurs through interactions between atoms and does not require the assistance of a nonlinear medium. Matter-wave FWM has attracted increasing attention due to its potential applications in matter-wave amplification \cite{PhysRevLett.95.170404, PhysRevA.71.041602, PhysRevLett.96.020406, PhysRevA.96.013605, PhysRevA.68.015603}, correlated particle pair generation \cite{PhysRevLett.89.020401, PhysRevLett.99.150405, PhysRevA.79.011601, PhysRevLett.105.190402}, and other nonclassical states of matter waves \cite{rolston2002nonlinear, molmer2003quantum, PhysRevLett.104.150402, goldstein1999quantum}, which will open up new possibilities for quantum information processing and precision measurement \cite{berrada2013integrated, bongs2004physics, liu2019interference, pan2019orbital, robins2013atom}.
	
So far, matter-wave FWM has been extensively studied, mainly focusing on two configurations. In the first configuration, all four participating momenta belong to the same spin component, leading to coherent scattering purely between momentum modes \cite{deng1999, PhysRevLett.89.020401, PhysRevA.68.015603, PhysRevA.96.013605, PhysRevLett.96.020406, PhysRevLett.99.150405, PhysRevLett.107.075301, OE.3.000530, PhysRevA.61.043604}. Deng \textit{et al}. experimentally confirmed for the first time that matter-wave FWM can occur in single-component Bose-Einstein condensates (BECs), which greatly stimulated the development of nonlinear atomic optics \cite{deng1999}. It has been shown that atomic FWM with large gain can generate two macroscopically occupied pair-correlated atomic beams  \cite{PhysRevLett.89.020401}. The degenerate FWM of matter waves was also observed in optical lattices \cite{PhysRevLett.95.170404, PhysRevA.71.041602, PhysRevLett.96.020406}. In the second configuration, the four participating momenta are distributed across different spin components, giving rise to coherent scattering that involves both momentum and spin degrees of freedom \cite{PhysRevA.70.033606, PhysRevLett.104.200402, PhysRevA.109.063316, hung2020four, PhysRevLett.107.210406}. Collinear matter-wave FWM has been observed in collisions between atoms in two spin states \cite{PhysRevLett.104.200402}, and has been shown to give rise to a periodic collapse and revival of the FWM yield when introducing spin exchange interactions \cite{PhysRevA.109.063316}. While atomic interactions are crucial to the FWM process, their tunable effects have not been systematically examined or compared in previous experiments.
	
\begin{figure*}[thb]
\includegraphics[scale=1.0]{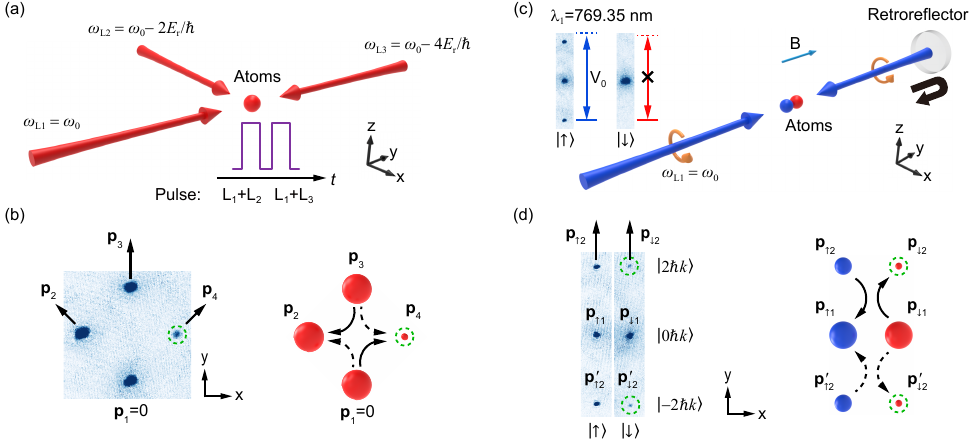}
\caption{\label{Fig1} (a) Experimental optical configuration for FWM of a single-spin component. Two sets of Bragg pulses $L_1+L_2$ and $L_1+L_3$ are applied sequentially to generate three distinct momentum states $\mathbf{p}_1$, $\mathbf{p}_2$, and $\mathbf{p}_3$, as shown in (b). The three light beams have nearly identical wavelength of $790.00\ \mathrm{nm}$. (b) The left panel shows a typical absorption image of a single-spin component FWM in a square geometric configuration, where the collision of three wave packets leads to the generation of a fourth wave packet. The right panel shows the corresponding schematic diagram of the FWM processes in which the solid arrows and dashed arrows represent two separated processes of atomic momentum transition. (c) Experimental optical configuration for FWM of two-spin components. The laser for the lattice beams operates at the tune-out wavelength of $769.35\ \mathrm{nm}$, and the bias magnetic field is oriented parallel to the laser propagation direction, which ensures that atoms in state $\lvert\uparrow\rangle$ experience the lattice potential $V_0$, while those in state $\lvert\downarrow\rangle$ are unaffected. (d) The left panel shows a typical absorption image of a two-spin component FWM in a collinear geometric configuration. When the initial atomic spin state is a mixture of $\lvert\uparrow\rangle$ and $\lvert\downarrow\rangle$, the $\lvert\downarrow\rangle$ state exhibits two scattered momentum states at $\lvert\pm 2 \hbar k\rangle$. The right panel shows the corresponding schematic diagram of the FWM processes in which the solid arrows and dashed arrows represent two symmetric FWM processes, respectively. In (b, d), the green dashed circles denote the wave packets generated by FWM.}
\label{fig1}
\end{figure*}

In this paper, we experimentally investigate matter-wave FWM in $^{39}$K BECs, which feature multiple Feshbach resonances that allow precise control over atomic interactions \cite{frolian2022realizing, semeghini2015measurement, etrych2023pinpointing, jorgensen2016observation}. By tuning the intra and interspin interactions between the two spin states, the system can be driven from the gas phase to the droplet phase. In the droplet regime, the mean-field effects become relatively weak, while quantum fluctuation effects described by the Lee-Huang-Yang correction manifest, potentially giving rise to interesting strongly correlated phenomena \cite{cabrera2018quantum, semeghini2018self, guo2021lee, cavicchioli2025dynamical, skov2021observation, d2019observation, cheiney2018bright, petrov2015quantum}. We investigate the influence of tunable atomic interaction on FWM in two configurations (see Fig. \hyperref[fig1]{1}) and find that the interaction significantly alters the FWM yield. In the configuration with single-spin components, the FWM yield increases with interaction strength. Particularly in the configuration with two-spin components, we perform FWM studies across both droplet and gas parameter regimes, observing the maximum FWM yield close to the critical region.
	
\section{EXPERIMENTAL RESULTS AND ANALYSIS}
\subsection{Experimental methods}

The experimental setup for producing the $^{39}$K BECs in various spin states has been detailed in our previous paper \cite{mi2021production}, so here we briefly introduce the main experimental process that should be noted in this work. Following the sympathetic cooling with $^{87}$Rb atoms, we successfully obtain dual-species BECs of $^{39}$K-$^{87}$Rb in a magnetic field of $120.16\ \mathrm{G}$, where both $^{87}$Rb and $^{39}$K atoms are in the $|F,m_F\rangle=\lvert1,-1\rangle$ state, with scattering lengths $a_{\text{KK}} = 13.2\ a_0$, $a_{\text{RbK}} = 58.1\ a_0$, and $a_{\text{RbRb}} = 100.4\ a_0$. Subsequently, by further reducing the depth of the optical dipole trap, the heavier $^{87}$Rb atoms are selectively expelled from the trap due to their greater mass compared to $^{39}$K atoms, and we eventually obtain single $^{39}$K BECs, with a typical temperature of $30(5)\  \mathrm{nK} $ and an atomic number of $3.0(2) \times 10^5$.

\begin{figure}[thb]
\includegraphics[scale=1.0]{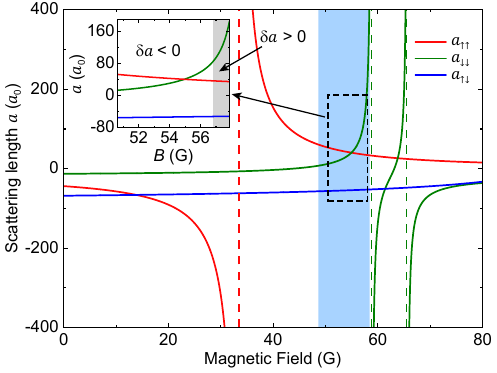}
\caption{\label{Fig2} Intraspin and interspin scattering lengths of $^{39}$K atoms as a function of Feshbach magnetic field, with the spin $\lvert\uparrow\rangle$ and $\lvert\downarrow\rangle$ labeling the $\lvert1,-1\rangle$ and $\lvert1,0\rangle$ states, respectively. The light blue shaded region denotes the magnetic field range employed for FWM in single-spin BECs. The top-left inset, a magnification of the dashed rectangle, denotes the magnetic field range employed for FWM in two-spin BECs, where the shaded region represents the gas regime with $\delta a > 0$ and the remaining region designates the droplet regime with $\delta a < 0$. }
\label{fig2}
\end{figure}

For the case of FWM in single-spin BECs, we transfer all atoms from the $\lvert1,-1\rangle$ (labeled as $\lvert\uparrow\rangle$) state to the $\lvert1,0\rangle$ (labeled as $\lvert\downarrow\rangle$) state at the magnetic field $57.51\ \mathrm{G}$. In the vicinity of Feshbach resonance with magnetic field $58.86\ \mathrm{G}$, the intraspin scattering length $a_{\downarrow\downarrow}$ can be well tuned from $7.4\ a_0$ to $485.1\ a_0$ in our experiment, as illustrated in Fig. \hyperref[fig2]{2}. For the case of FWM in two-spin BECs, we prepare the atoms in mixed spin states $\lvert\uparrow\rangle$ and $\lvert\downarrow\rangle$ with a ratio of $1:1$. We tune the magnetic field from $50.65\ \mathrm{G}$ to $57.87\ \mathrm{G}$, during which the intraspin scattering length $a_{\downarrow\downarrow}$ increases from $12.7\ a_0$ to $185.9\ a_0$, $a_{\uparrow\uparrow}$ varies from $52.9\ a_0$ to $31.4\ a_0$, and the interspin scattering length $a_{\uparrow\downarrow}$ remains nearly constant, as shown in the inset of Fig. \hyperref[fig2]{2}. The scattering length plays a crucial role in determining the phase diagram between the gas and droplet regimes. The phase boundary is characterized by $\delta a = a_{\uparrow\downarrow} + \sqrt{a_{\uparrow\uparrow} a_{\downarrow\downarrow}}$, which represents the residual mean-field interaction. When $\delta a > 0$, the system remains in the gas phase, whereas $\delta a < 0$ corresponds to the droplet regime \cite{cabrera2018quantum}.

\subsection{Matter-wave FWM in single-spin BECs}

We first perform the FWM experiment in single-spin BECs. For a given scattering length set by the magnetic field, atoms initially in the zero-momentum state $\mathbf{p}_1 = 0$ are partially transferred by two sequential Bragg pulses into two additional momentum states, $\mathbf{p}_2 = -\hbar k \mathbf{e}_x + \hbar k \mathbf{e}_y$ and $\mathbf{p}_3 = 2\hbar k \mathbf{e}_y$, where $\hbar$ is the reduced Planck constant, $k=2\pi/\lambda$ is the wave vector of the laser, and $\lambda = 790.00\ \mathrm{nm}$. The optical configuration of the Bragg pulses is illustrated in Fig. \hyperref[fig1]{1(a)}, where the three laser beams share almost the same wavelength, but have specific frequency differences for resonant Bragg scattering. The frequency difference between $L_1$ and $L_3$ is set to $4E_\mathrm{r}$, and the frequency difference between $L_1$ and $L_2$ is set to $2E_\mathrm{r}$, with $E_{\mathrm{r}} = \hbar^2 k^2/(2m) = h \times 8.2\ \mathrm{kHz}$ being the photon recoil energy, which ensures that the energy differences of the scattered photons compensate for the kinetic energy changes of atoms transitioning between different momentum states. The relative population of atoms across the three momenta is well controlled by adjusting the Bragg pulse durations. In the experiments, both pulses have a width of 24\ $\mu$s with a 16\ $\mu$s interval, resulting in an equal distribution of atoms among the three momenta.

The collisions between atoms from the three momentum modes generate a new momentum mode $\mathbf{p}_4 = \hbar k \mathbf{e}_x + \hbar k \mathbf{e}_y$. Such a matter-wave FWM process satisfies both momentum conservation ($\mathbf{p}_1 + \mathbf{p}_3 = \mathbf{p}_2 + \mathbf{p}_4$) and energy conservation ($\mathbf{p}_1^2 + \mathbf{p}_3^2 = \mathbf{p}_2^2 + \mathbf{p}_4^2$). In the experiment, to remove the influence of the optical dipole trap, we first prepare the BECs in three momentum states and then switch off the optical dipole trap, allowing the atoms to evolve in free space. The atomic momentum distribution after the FWM process is detected by time-of-flight (TOF) absorption imaging. The atomic cloud first expands for $5\ \mathrm{ms}$ under the high magnetic field of the FWM, which releases its internal energy. We then adjust the bias magnetic field to $3.7\ \mathrm{G}$ and let the cloud undergo an additional $15\ \mathrm{ms}$ of free flight. Finally, a probe beam is applied to perform the absorption imaging. Figure \hyperref[fig1]{1(b)} shows a typical absorption imaging picture, in which we can clearly see that a new momentum mode $\mathbf{p}_4$ has been generated.

\begin{figure}[tb]
\includegraphics[scale=1.0]{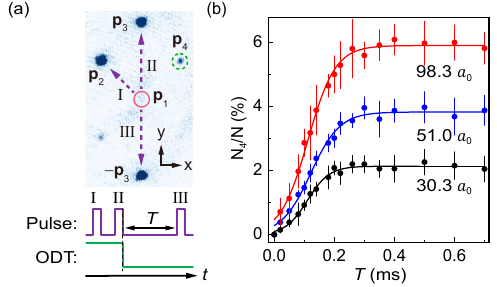}
\caption{\label{Fig3} (a) Schematic illustration of interrupting the FWM process. Pulses I and II prepare three momentum states for FWM, while pulse III (30\ $\mu$s) removes atoms from momentum state $\mathbf{p}_1$ to $-\mathbf{p}_3$, thereby interrupting the process. The optical dipole trap (ODT) is turned off after pulse I and II, to allow the wave packets to evolve in free space for a duration of $T$. (b) The growth curves of the FWM processes with different scattering lengths. The solid lines represent sigmoidal fits to the experimental data. The fitting formula is $f(T) = A / \{1 + \exp[-K(T - T_c)]\}$, where $A$ represents the maximum value of the curve, $T_c$ denotes the time at half-maximum, and $K$ indicates the growth rate. The fitted $T_c$ values are $0.11\ \mathrm{ms}$, $0.12\ \mathrm{ms}$, and $0.11\ \mathrm{ms}$, respectively. Each data point represents an average over five experimental runs, and the error bars indicate standard deviations of repeated experimental measurements.}
\label{fig3}
\end{figure}

The FWM evolution process can be terminated by applying a third Bragg pulse. This pulse utilizes the beam $L_1$ and $L_3$, with the frequency of $L_3$ set to $\omega_0 + 4E_{\mathrm{r}} / \hbar$, where $\omega_0$ is the frequency of $L_1$. Such a frequency setting enables the transfer of atoms from momentum $\mathbf{p}_1$ to $-\mathbf{p}_3$. The absorption imaging picture shown in the upper panel of Fig. \hyperref[fig3]{3(a)} demonstrates that the third Bragg pulse effectively transfers nearly all atoms from the $\mathbf{p}_1$ state to the $-\mathbf{p}_3$ state, while a small fraction of atoms in the $\mathbf{p}_2$ and $\mathbf{p}_4$ states undergo nonresonant Bragg scattering to the $-\mathbf{p}_4$ and $-\mathbf{p}_2$ states, respectively. The new configuration of three momenta ($-\mathbf{p}_3$, $\mathbf{p}_2$, $\mathbf{p}_3$) can no longer generate a new momentum mode $\mathbf{p}_4'$ that satisfies both momentum and energy conservation, thereby terminating the FWM process.

The FWM process can be measured by interrupting it via the third Bragg pulse following an evolution time of $T$. The evolution process of FWM under varying scattering lengths $a_{\downarrow\downarrow}$ is shown in Fig. \hyperref[fig3]{3(b)}, where we systematically set $a_{\downarrow\downarrow}$ to $30.3\ a_0$, $51.0\ a_0$, and $98.3\ a_0$, respectively. We find that as the scattering length increases, the generated FWM mode grows faster and eventually saturates at a higher population. Notably, all growth curves in Fig. \hyperref[fig3]{3(b)} reach their maxima at approximately $T = 0.3\ \mathrm{ms}$, indicating that the three wave packets no longer perfectly overlap with each other at this moment.

\begin{figure}[tb]
\includegraphics[scale=1.0]{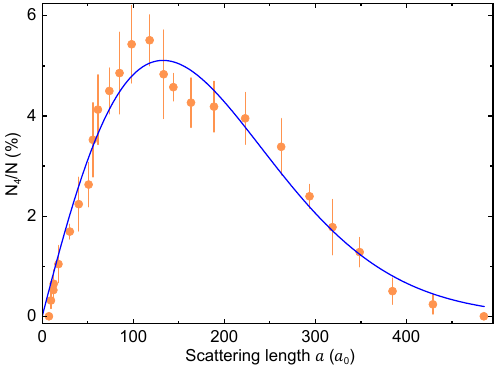}
\caption{\label{Fig4} The FWM yield of the single-spin components under various scattering lengths. The FWM yield vanishes as the scattering length approaches zero, reaches its maximum in the scattering length around $118\ a_0$, and then decreases with further increased scattering length due to enhanced three-body losses. Each data point represents an average over five experimental runs, and the error bars indicate standard deviations of repeated experimental measurements. The blue solid line presents the fit of experimental data based on an empirical formula $f(x) = ax\exp(-x^b / c)$. Corresponding to the units of the horizontal and vertical axes in the figure, we obtain the dimensionless fitting parameters as $a = 7.0435 \times 10^{-4}$, $b = 1.6550$, and $c = 5401.0$.}
\label{fig4}
\end{figure}

We also vary the scattering length $a_{\downarrow\downarrow}$ over a wider range and measure the corresponding FWM yield, as shown in Fig. \hyperref[fig4]{4}. When $a_{\downarrow\downarrow}$ approaches zero, almost no atoms are scattered into the $\mathbf{p}_4$ momentum state. As $a_{\downarrow\downarrow}$ increases, the FWM yield rises accordingly and reaches a maximum of $5.5\%$ near $118\ a_0$. With the further increase of $a_{\downarrow\downarrow}$, the FWM yield decreases and eventually approaches zero. The experimental data can be well fitted by an empirical formula $f(x) = ax\exp(-x^b / c)$. On the right side of the formula, the first $x$ comes from the coulpling strength of the FWM process, which varies linearly with the scattering length $a_{\downarrow\downarrow}$ \cite{PhysRevA.70.033606, PhysRevA.62.023608}. The exponential dependence can be attributed to several mechanisms, including three-body loss, spontaneous \textit{s}-wave scattering between atoms with different momenta, and the quantum depletion effect. As the scattering length increases, each of these mechanisms exerts a more pronounced influence, ultimately resulting in atomic loss as well as a reduction in the condensate fraction within ultracold atomic gases, which in turn lowers the FWM yield.
	
\subsection{Matter-wave FWM in two-spin BECs}	
Next, we investigate the FWM process in BECs of atoms with two spin states. The introduction of spin degree of freedom into FWM enables more flexible momentum configurations. Unlike the square momentum configuration in the single-component systems, FWM in two-component BECs can also be realized in a collinear momentum configuration. We utilize the tune-out wavelength of the $\lvert\downarrow\rangle$ state to create a one-dimensional spin-dependent optical lattice. In this configuration, the $\lvert\uparrow\rangle$ state is subjected to the lattice potential, whereas the $\lvert\downarrow\rangle$ state remains unaffected by any optical lattice potential \cite{wen2021experimental, mandel2003coherent, yang2017spin}. In this lattice, atoms in the $\lvert\uparrow\rangle$ state can be transferred to high-momentum states via Kapitza-Dirac (KD) scattering, whereas $\lvert\downarrow\rangle$ atoms remain in the zero-momentum state. By precise wavelength scanning, we determine the tune-out wavelength to be $769.35\ \mathrm{nm}$. Figure \hyperref[fig1]{1(c)} shows that at this wavelength, when all atoms occupy the $\lvert\uparrow\rangle$ state, the KD scattering pulse efficiently transfers them to the $\lvert\pm 2\hbar k\rangle$ momentum states. In contrast, for atoms in the $\lvert\downarrow\rangle$ state, the lattice pulse induces no momentum transfer. The population transfer fraction of $\lvert\uparrow\rangle$ atoms to higher momentum states can be controlled by adjusting the optical lattice potential depth.

Interestingly, when atoms are prepared in a mixture of $\lvert\uparrow\rangle$ and $\lvert\downarrow\rangle$ states, in addition to the initial momentum states prepared by the spin-dependent optical lattice, two new high momentum states emerge in the $\lvert\downarrow\rangle$ component, as illustrated in Fig. \hyperref[fig1]{1(d)}. The generation of new momentum states is attributed to two symmetrical and simultaneous FWM processes, as indicated respectively by the solid and dashed arrows in the right panel of Fig. \hyperref[fig1]{1(d)}. Both processes satisfy momentum conservation (such as $\mathbf{p}_{\downarrow 1} + \mathbf{p}_{\uparrow 2} = \mathbf{p}_{\downarrow 2} + \mathbf{p}_{\uparrow 1}$) and energy conservation ($\mathbf{p}^2_{\downarrow 1} + \mathbf{p}^2_{\uparrow 2} = \mathbf{p}^2_{\downarrow 2} + \mathbf{p}^2_{\uparrow 1}$).

In our experiments, the BECs are initially prepared in the $\lvert\uparrow\rangle$ state with the magnetic field stabilized at the target point. Then, approximately half of the atoms are transferred to the $\lvert\downarrow\rangle$ state within 1\ ms via a radiofrequency pulse. Immediately after, a 12\ $\mu$s KD scattering pulse with a lattice depth of $V_0 = 6 E_\mathrm{r}$ (for atoms in $\lvert\uparrow\rangle$ state) is applied to prepare the initial momentum modes for FWM. The optical trap is then switched off to start FWM evolution in free space. We employ the same TOF absorption imaging procedure as in the single-component FWM to observe the momentum distribution. To distinguish the spin states of atoms in the same momentum states, we apply a Stern-Gerlach gradient magnetic field along the horizontal $x$ direction during the final $15\ \mathrm{ms}$ of the TOF period.
	
\begin{figure}[tb]
\includegraphics[scale=1.0]{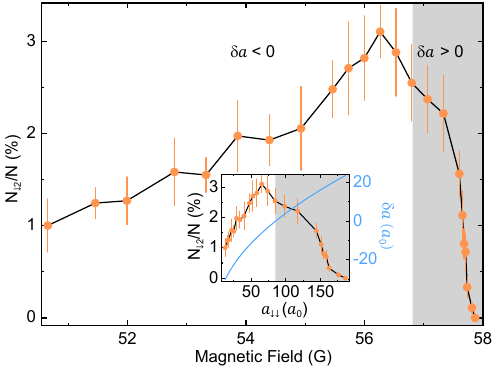}
\caption{\label{Fig5} The FWM yield of two-spin components under various Feshbach magnetic field strengths. The inset shows the curves of the FWM yield and $\delta a$ versus the scattering length $a_{\downarrow\downarrow}$. The FWM yield reaches a maximum around $\delta a = -6\ a_0$. Each data point represents an average over five experimental runs, and the error bars indicate standard deviations of repeated experimental measurements.}
\label{fig5}
\end{figure}
	
We measure the influence of the atomic scattering lengths on the yield of FWM, as shown in Fig. \hyperref[fig5]{5}. As discussed previously, upon adjusting the magnetic-field strength, the mixture of two spin states traverses two distinct parameter regimes corresponding to the gas phase with the scattering lengths satisfying $\delta a > 0$ and the quantum droplet phase with $\delta a < 0$. It is found that the FWM yield reaches its maximum in the droplet parameter regime near the critical region, and gradually decreases when moving the system away from this region. This can be understood by noting that within the droplet phase regime, the wave packets maintain a relatively high atomic number density throughout the FWM evolution process, thereby achieving maximum FWM yield.

\begin{figure}[tb]
\includegraphics[scale=1.0]{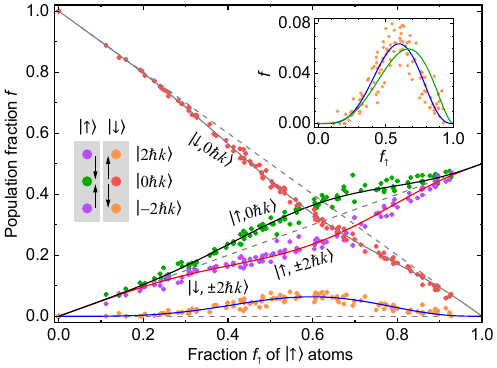}
\caption{\label{Fig6} Population fractions of all momentum modes involved in FWM as a function of the population fraction in the $\lvert\uparrow\rangle$ state. Each data point presents a single experimental run. The blue solid line presents the fitting of the population fraction in $\lvert\downarrow,\pm2\hbar k\rangle$ mode based on an empirical formula $f(x) = ax^{2+2b}(1-x)^{1+c}$. The dimensionless fitting parameters are $a = 10.974$, $b = 1.2829$, and $c = 2.0723$. These parameters are then used with appropriately adapted formulas (see main text) to fit the population fractions in the remaining momentum modes, represented by the other colored solid lines. All the fitting lines agree well with the experimental data points, indicating that the population transfer among all momentum modes is highly correlated. The dashed lines indicate the initial population fractions before FWM. The arrows in the legend indicate the population transfer between different momentum modes. The inset shows two fitting lines of the population fraction in the $\lvert\downarrow,\pm2\hbar k\rangle$ mode. The blue solid line is identical to that in the main figure, whereas the green solid line represents a fit using the formula $f(x) = ax^{2+2b}(1-x)^{1+b}$, with fitted parameter values $a = 1.8343$ and $b = 0.79350$. The former fit accounts for the difference in intraspin scattering lengths between the $\lvert\uparrow\rangle$ and $\lvert\downarrow\rangle$ spin states, while the latter does not. 
The blue solid line provides a better fit to the experimental data and exhibits a peak at a lower value of fraction $f_{\downarrow}$.}
\label{fig6}
\end{figure}
	
The atomic population among the two spin states significantly influences the FWM yield. In the experiments, the Feshbach magnetic field is set at $B = 56.79\ \mathrm{G}$, which is close to the value corresponding to the maximum FWM yield in Fig. \hyperref[fig5]{5}. Here we have $a_{\uparrow\downarrow} = -53.2\ a_0$, $a_{\uparrow\uparrow} = 33.3\ a_0$, $a_{\downarrow\downarrow} = 84.6\ a_0$, and $\delta a = -0.2\ a_0$. Then we vary the population fraction between the two spin states and measure the atomic distributions of all momentum states after the FWM evolution, as shown in Fig. \hyperref[fig6]{6}. From Fig. \hyperref[fig6]{6}, one can find that under FWM the atom numbers at the momentum states $\lvert \uparrow,\pm 2\hbar k \rangle$ and $\lvert \downarrow,0 \rangle$ decrease, while those at $\lvert \downarrow,\pm 2\hbar k \rangle$ and $\lvert \uparrow,0 \rangle$ increase. These population changes induced by FWM in different momentum modes are consistent with the process illustrated in Fig. \hyperref[fig1]{1(d)}  (also depicted in the legend of Fig. \hyperref[fig6]{6}). 
 
The experimental data presented in Fig. \hyperref[fig6]{6} can be interpreted through the following simplified physical picture. The growth rate of FWM is governed by the coupling term $U_{\uparrow\downarrow}\Phi_{\uparrow 1}^* \Phi_{\uparrow 2}\Phi_{\downarrow 1}$ and $U_{\uparrow\downarrow}\Phi_{\uparrow 1}^* \Phi'_{\uparrow 2}\Phi_{\downarrow 1}$ for the upper and lower FWM processes shown in Fig. \hyperref[fig1]{1(d)}, respectively \cite{PhysRevA.70.033606, PhysRevA.62.023608, PhysRevLett.104.200402, PhysRevA.109.063316}. Here $U_{\uparrow\downarrow}=4\pi\hbar^2 a_{\uparrow\downarrow}/m$, $m$ is the mass of colliding atoms and $\Phi$ is the wave function of corresponding momentum mode. Among the three scattering lengths of the atomic mixture,  only the interspin scattering length $a_{\uparrow\downarrow}$ directly determines the coupling strength of FWM. When the atomic population distribution among different momentum modes is adjusted, both the growth rate and the final FWM yield vary accordingly. Denoting the fraction of total atoms in $\lvert\uparrow\rangle$ state before FWM evolution as $x$, then the population fraction in $\lvert\downarrow,0\hbar k\rangle$ mode should be $1-x$. The KD pulse applied in the experiments fixes the fraction in $\lvert\uparrow,0\hbar k\rangle$ mode at $x/2$, leaving $x/4$ each in the $\lvert\uparrow,2\hbar k\rangle$ and $\lvert\uparrow,-2\hbar k\rangle$ modes. Thus, the FWM yield is expected to scale as $x^2(1-x)$. Additionally, the size of the matter-wave packet increases with atomic number, and a larger wave-packet size allows for a longer evolution time of FWM, thereby increasing the FWM yield. Incorporating the dependence of wave-packet size on atomic number, we establish the empirical formula as $f(x) = ax^{2+2b}(1-x)^{1+c}$. As illustrated by the blue solid line in Fig. \hyperref[fig6]{6}, this formula provides a nice fit for the population fraction in $\lvert\downarrow,\pm2\hbar k\rangle$ mode. 

From the FWM process shown in Fig. \hyperref[fig1]{1(d)}, it can be seen that the population transfer of the atoms among different momentum modes is highly correlated. Therefore, we can infer that after FWM, the population fractions of atoms in the $\lvert\downarrow,0\hbar k\rangle$, $\lvert\uparrow,0\hbar k\rangle$, and $\lvert\uparrow,\pm2\hbar k\rangle$ modes are $g(x) = 1 - x - f(x)$, $h(x) = x/2 + f(x)$, and $q(x) = x/2 - f(x)$, respectively. As indicated by the colored solid lines in Fig. \hyperref[fig6]{6}, using the same fitting parameters $a$, $b$, and $c$ of $f(x)$, we can well fit the population fractions in all the other momentum modes by $g(x)$, $h(x)$, and $q(x)$. When fitting the experimental data, we find that different parameters $b$ and $c$ need to be introduced for the wave packets in $\lvert\uparrow\rangle$ and $\lvert\downarrow\rangle$ states, primarily due to the different intraspin scattering lengths of atoms in these two spin states. The inset in Fig. \hyperref[fig6]{6} compares the fitting results for $c \neq b$ and $c = b$, with the former showing better agreement with the experimental data. The former fitting gives the fitting parameters $c > b$, which is reasonable since we have $a_{\downarrow\downarrow} > a_{\uparrow\uparrow}$ in our experiments. When $c = b$ is enforced, the function $f(x)$ reaches its maximum at $x = 2/3$; however, experimentally, the maximum FWM yield occurs near $x = 0.6$. The fitting with $c > b$ yields a more accurate peak position.

\section{CONCLUSION}
In summary, we investigate the FWM process with tunable atomic interaction in both single-component and two-component BECs of  $^{39}$K atoms under two distinct momentum configurations. By precisely tuning the atomic interaction via the Feshbach resonance in both configurations, the FWM yield can be significantly altered. In the square configuration of single-component BECs, the experimentally tunable scattering length directly determines the coupling strength among matter-wave packets. Consequently, the FWM yield initially increases with the scattering length. As the scattering length increases further, however, effects such as three-body loss, spontaneous $s$-wave scattering between colliding atomic clouds, and quantum depletion become significantly enhanced. These effects substantially reduce both the atom number and the condensate fraction in the ultracold atomic gases, ultimately causing the FWM yield to decline. In the collinear configuration of the two-component system, the tunable scattering length mainly influences the size and spatial overlap of the wave packets. We find that the FWM yield reaches its maximum in the droplet parameter regime close to the critical region between the gas and the droplet phases. The quantum fluctuations in the droplet regime make the investigation of FWM particularly compelling. This effect not only paves the way for potential applications in quantum information processing and precision measurement due to a significantly increased FWM yield, but also offers a platform for exploring novel quantum many-body dynamics.

\begin{acknowledgments}
		
This research is supported by the National Key Research and Development Program of China (Grants No. 2021YFA1401700 and No. 2022YFA1404101), the National Natural Science Foundation of China (Grants No. 12034011, No. U23A6004, No. 12488301, No. 92476001, No. 12374245, No. 12474266, No. 12474252, No. 12322409, and No. 12004229), and the Innovation Program for Quantum Science and Technology (Grant No. 2021ZD0302003).
		
\end{acknowledgments}
	
\bibliography{references}
	
\end{document}